\RequirePackage{fix-cm}
\documentclass[twocolumn]{svjour3}          
\smartqed  
\usepackage{graphicx}
\graphicspath{ {Figures/} }
\usepackage{algorithm}
\usepackage[usenames, dvipsnames]{color}
\usepackage[noend]{algpseudocode}
\usepackage{url}
\usepackage{subfigure}
\usepackage{amsmath}
\usepackage{blindtext}
\usepackage{enumerate}
\usepackage[table,xcdraw]{xcolor}
\usepackage{tabularx}

\newcommand{\para}[1]{\medskip \noindent {\bf #1}}

\makeatletter
\newcommand\footnoteref[1]{\protected@xdef\@thefnmark{\ref{#1}}\@footnotemark}
\makeatother

\newcommand{\x}{\mathbf{x}}
\newcommand{\y}{\mathbf{y}}
\renewcommand{\u}{\mathbf{u}}
\newcommand{\N}{\mathbf{\mathcal{N}}}

\begin{document}

\title{A software-defined architecture for control of IoT Cyberphysical Systems
}


\author{Ala' Darabseh$^1$         \and
       Nikolaos M. Freris$^2$
       }


\institute{
              New York University Abu Dhabi, UAE, P.O Box: 129188              
           \and
           \\          
           Tel.: +971-26284649$^1$,+971-26284823$^2$ \\
           Fax: +971-26284000 \\          
           \email{afd8@nyu.edu}$^1$, \email{nf47@nyu.edu}$^2$
           }

\date{}

\maketitle

\begin{abstract}
	
Based on software-defined principles, we propose a holistic architecture for \emph{Cyberphysical Systems} (CPS) and \emph{Internet of Things}  (IoT) applications,  and highlight the merits pertaining to scalability, flexibility, robustness, interoperability, and cyber security. Our design especially capitalizes on the computational units possessed by smart agents, which may be utilized for decentralized control and in-network data processing.  
We characterize the data flow, communication flow, and control flow that 
assimilate a set of components such as sensors, actuators, controllers, and coordinators in a systemic programmable fashion. We specifically aim for distributed and decentralized decision-making by spreading the control over several hierarchical layers. 
In addition, we propose a middleware layer to encapsulate units and services for time-critical operations in highly dynamic environments. 
We further enlist a multitude of vulnerabilities to cyberattacks, and integrate software-defined solutions for enabling resilience, detection and recovery. In this purview, several controllers cooperate to identify and respond to security threats and abnormal situations in a    
self-adjusting manner. Last, we illustrate numerical simulations in support of the virtues of a software-defined design for CPS and IoT.  
	
\end{abstract}

\keywords {Software Defined Systems (SDSys) \and Internet of Things (IoT) \and Cyberphysical Systems (CPS) \and Distributed Systems \and Decentralized Control \and Cyber Security \and Middleware}

\section*{Acknowledgment}
This work was supported in part by the Center for Cyber Security at New York University Abu Dhabi.
\section{Introduction}
\label{sec:intro}

The Internet of Things (IoT)~\cite{gubbi2013internet} interconnects a huge number of `smart' devices (such as mobile phones, sensors, routers, microcontrollers) alongside large data centers, and provides mechanisms for collection and processing of big data, communication, as well as cloud services. A closely related framework lies in 
\emph{Cyberphysical Systems} (CPS)~\cite{rajkumar2010cyber,kim2012cyber}, 
where smart agents that possess sensing, computation, communication, and control capabilities are internetworked to control physical entities and processes. 
Prominent applications enlist Intelligent Transportation Systems (ITS), smart grids,  wireless sensor networks~\cite{WSN}, smart buildings, and mobile healthcare  (mHealth).   

%

At such large scale, existing control architectures far exceed their capacity in  efficiently administering the conjugation of physical space with the cyberspace. 
Notable challenges that have to be accounted for in IoT and CPS applications feature the urge for adaptability, scalability, security, safety, and robustness to abrupt changes in the modus operandi of the network~\cite{rajkumar2010cyber,lee2008cyber}. A key pathway is to design \emph{hybrid systems}~\cite{hybrid}, in which the software controls are decoupled from the embedded components~\cite{lee2011introduction}. 

   
Software Defined Systems (SDSys) come a systematic paradigm to design such systems by abstracting the controls laws from the hardware devices at the physical layer, and placing them in a software-defined control layer. Such decoupling is intended to provide reliable, cost-effective, and real-time control solutions for CPS and IoT. This concept extends and expands structured development of large-scale software, and 
was first introduced within the context of cognitive radios~\cite{mitola2000cognitive}. Subsequently,  it has been used to develop Software Defined Networking (SDN)~\cite{jain2013network} as well as in several facets and aspects of IoT~\cite{SDIoTAla,al2016novel,SDCloud}.


In this paper, we utilize software defined principles to propose a comprehensive architectural design for CPS and IoT systems. 
The proposed architecture specifically intends for decentralized decision-making within the IoT, by leveraging the computational resources that are ubiquitous within the many entities it comprises. We specify how the proposed model reduces the control complexity and allows for flexible integration and adaptation in the cyberspace. 
Our architecture entails three main domains: 
the physical space, the cyberspace, and the structured control space, all being SDSys-described. The hierarchical and decentralized structure of the control space is carefully designed in a way that assigns the responsibilities of each agent within the IoT technology domain. The services of the middleware layer ~\cite{kim2013real} of the proposed model are abstracted and enhanced to pledge increased safety, reliability and performance in a highly dynamic environment, primarily due to changes in the network topology (mainly resulting from agent mobility and battery drainage in portable devices). 
Moreover, we specify key requirements and software-defined solutions for achieving high quality-of-service (QoS) alongside cyber security. To this end, we interplay both a bottom-up and a top-bottom workflow for spreading information and actuation throughout the network. In addition, we identify several classes of potential cyber attacks and vulnerabilities across the multiple levels, and propose effective detection and recovery solutions.     
Finally, we have built an object-oriented simulator in Python using features and principles from general purpose SDSys simulators such as Mininet, Maxinet and Mininet-WiFi, and use it to test and evaluate several performance indicators of our proposed modules. 
%
%
%
%
%

This paper extends and expands our preliminary studies in~\cite{SDCPSconf} in several directions: a) we provide a more detailed description of several important attributes of the architecture and solidify the connections with IoT technology and fledgling applications; b) we explicitly discuss a software-defined design for cyber security in CPS and IoT applications; c) we devise an object-oriented simulator and verify the merits of our approach via various simulations studies.

%
%
%

The remainder of the paper is structured as follows.
We discuss the main design challenges pertinent to CPS in Sec.~\ref{sec:CPS}, and recap the key concepts of Software Defined Systems (SDSys) in Sec.~\ref{sec:sdsys}.  Sec.~\ref{sec:SystemModel} is designated for the proposed software-defined CPS architecture (SDCPS): we expose the model requirements (Sec.~\ref{sec:requirements}),  the architectural overview (Sec.~\ref{sec:overview}), the main elements (Sec.~\ref{sec:elements}), the control architecture (Sec. \ref{sec:control}), the middleware layer (Sec.~\ref{sec:middleware}), the work-flow (Sec. \ref{sec:flow}) and other important features (Sec.~\ref{sec:features}). We test aspects of our proposed solution in Sec.~\ref{sec:exp}. 
Sec.~\ref{sec:con} concludes the paper.
    
\subsection {Design challenges in CPS}
\label{sec:CPS}

In order for CPS to truly emerge as the fourth industrial revolution they envisioned~\cite{rajkumar2010cyber}, various attributes in modeling, communication, sensing, control, and cyber security have to be attended to.  
We expose a brief overview of the main challenges and design requirements in this fascinating domain~\cite{gungor2009industrial}:

\para{Large volumes of data:} Sensors constitute an effective bridge between the cyberspace and the physical space. In large-scale CPS, such as transportation and sensor networks~\cite{WSN}, real-time sensing produces really \emph{big data}. It is indispensable to provide algorithms for efficiently filtering and mining big data~\cite{distances,watermarking} in the real-time~\cite{RCS,EUSIPCO}. 
%

\para{Scalability:} The large number of devices in CPS that come equipped with heterogeneous hardware and software is a prior aspect to tackle in CPS ~\cite{rajkumar2010cyber}. 
It is integral to administer the right APIs for  off-the-shelf integration, and autonomous configuration, accompanied with new theories for scalable inference and control~\cite{rksimax1}.  

\para{Real-Time Decisions:} CPS operate subject to stringent real-time constraints
in communication, computation and control.  It is therefore crucial to account for deadlines in allocating resources and making decisions~\cite{hou2009theory}, to develop algorithms for online computing~\cite{RCS,EUSIPCO}, as well as software that can facilitate a smooth operation in the real-time~\cite{kim2013real}. 


\para{Security and Fault-Tolerance:}  The coupling between the cyber and the physical spaces open the door for many vulnerabilities to attacks and components failures. In the regime of CPS, there is a need for new algorithms that relax the implicit assumption of benign agents and fault-free operation, which can further provide theoretical guarantees in the presence of malevolent or Byzantine users~\cite{cardenas2008secure,SATS,DSC}.



\para{New Theories:}  In the light of overwhelming technological advances, arises the need for new theories~\cite{gupta2000capacity,freris2011fundamental,WSN,kim2012cyber} that approach CPS from a fundamental theoretical viewpoint and outcome efficient, low-complexity algorithms with provable performance guarantees~\cite{distances,rksimax1}. 

To summarize, modeling and algorithmic tools from system theory and software engineering associated with abstraction, wireless networking, system verification, control, and fault tolerance have to be invoked, but at the same time revisited. To this end, distributed, decentralized, and software-defined solutions can set the stage for building the engineering systems of the (not so remote) future.

\subsection{Software Defined Systems (SDSys)}
\label{sec:sdsys}

Software Defined Systems (SDSys) aim to decouple the physical space from the cyberspace by retracting controls from the embedded hardware and abstracting them into a software layer. There has been extensive work in several contexts: Radios (SDR)~\cite{mitola2000cognitive}, Networking (SDN) \cite{jain2013network}, Security~\cite{sdsec2}, Storage~\cite{sdstor2,SDCache} and Cloud~\cite{SDCloud}. 
Software-Defined Networking (SDN) defines a new way to control the process of forwarding packets in a network. 
In this context, open-source Python-based simulators are available to evaluate the performance of SDN-based protocols, such as 
%
Mininet~\cite{Mininet}, Mininet-WiFi~\cite{fontes2015mininet}, Maxinet~\cite{Maxinet}.  
%
%
%
%
%
A  software-defined model for cloud management which may effectively mitigate several cyber-threats was proposed in~\cite{SDCloud}. 

\section{SDCPS : The proposed architecture for CPS}
\label{sec:SystemModel}

In this section, we illustrate our proposed architecture for cyberphysical systems and IoT applications that 
which builds upon software-defined principles. 
The abundant and ubiquitous communication and computation power of smart devices is  exploited to introduce a light-weight, secure and reliable systemic control solution for real-time management of IoT systems.    

\subsection{Model Requirements}\label{sec:requirements}


We define a set of requirements and classify them in two main categories: 1) quality-of-service (QoS) requirements, and 2) security requirements.

\para{QoS Requirements:}

\begin{enumerate}
	
	\item \emph{Resource Exploitation:} Resources are plentiful in a CPS. An effective management mechanism is one that maximizes the benefits from exploiting as much as possible the available computational, communication, sensing, and actuation amenities in a coherent and concurrent manner. 
	\item \emph{Load Balancing:} It is important to leverage resources in a \emph{fair} manner. Service requests should be handled timely, while maintaining load balancing among the various system controllers. A key objective of the schedulers is to distribute the workload as evenly as possible so as to alleviate network congestion, minimize delays, and avoid bottlenecks (in the sense of communication or computation over-use) which may drain the batteries of remote smart devices and consequently drastically compromise system operation. 
%
	
	
	
	\item \emph{Real-time:} Controlling physical entities such as cars, sensors, smart grids and medicare operations imposes rigid real-time constraints. It is therefore important to define deadlines for the completion of time-critical tasks and devise real-time schedulers~\cite{hou2009theory,kim2013real} that provably honor them.
	
\end{enumerate}  
     
\para{Security Requirements:}


Cyber security constitutes a chief concern in modern networked-control systems, where  it is of vital importance to design and implement effective mechanisms to prevent,  detect and recover from a range of cyber-attacks. This objective can be accommodated in multiple complementary ways such as: a) \emph{encryption}, which yields both data privacy as well as security in packet-based communication~\cite{DSC}, b) control-based approaches \cite{tabuada,SATS,satchidanandan2017dynamic} that seek to identify attacks as well as to handle `stealthy' attacks (by assuring that undetectable attacks may not harm the system operation), c) software-based solutions, where different 
tasks and entities are dynamically assigned privileges by a dedicated security unit in the system.


In CPS, attack strategies are constantly evolving. Accordingly, the various tools for scanning, isolating, and resolving threats have to be constantly updated to meet their crucial mission. Several attack models apply in the regime of CPS and IoT systems (see also~\cite{yampolskiy2013taxonomy}):  


\begin{itemize}
	\item \emph{Trojans:} Trojans are executable program files injected by malicious users into the Internet. The effects of a trojan are triggered when downloaded and executed by a user. There are several types: Sending Trojans, Remote-access Trojans, Proxy Trojans, Denial-of-Service (DoS) Trojans and many others. 
	

\item \emph{DoS/DDoS attack:} Denial-of-Service (DoS) refers to the action of 
preventing a user from accessing system resources. Distributed Denial-of-Service (DDoS) is a type of DoS attack where multiple compromised systems are used to target a single system by rendering a subset of resources inaccessible to it. 

\item \emph{Packet Forging attack:} This is also known as ``packet injection'' and
entails creating seemingly normal packets to interrupt direct communication between users. Consequently, ``man-in-the-middle'' attacks can be launched, where an attacker secretly relays and possibly alters the messages between two parties under the perception of directly communicating with each other. There are several tools that an attacker can use to generate such attacks, e.g., TCPinject and packETH.
	
\item \emph{Fingerprinting attack:} 
An attacker eavesdrops the conversation (even when encrypted) between two users  and obtains possession of some critical features of the sender/receiver in order to identify the network status and analyze traffic patterns with the intention of deploying harmful actions.  
	
\item \emph{Application Layer attack:}  
 This genre is classified into four types: the first one uses the HTTP protocol requests to overwhelm a site, the second one reflects threats to the SMTP protocol, the third one to the FTP protocol and the last one concerns SNMP attacks intending to monitoring and reconfiguring the system.       
 Detecting such attacks is typically much harder than the detection of attacks on the network layer.

\item \emph{User attacks:} In this type of attack, a malicious user seeks to 
trick the supervisor into obtaining the same privileges as a legitimate user, e.g.,  by exploiting vulnerabilities in a local machine to create an account inside it. There are different forms of this attack, such as U2R and R2L attacks.  
		
\end{itemize}


Malevolent users exploit the heterogeneity of CPS to launch attacks to all system  components; this leads to the additional taxonomy of attacks into:

\begin{enumerate}
	\item \emph{Sensor-related attacks:} An attacker tries to eavesdrop or alter sensed data in order to compromise the system operation.
	\item \emph{Actuator-related attacks:} An attacker seeks to change the control commands of an actuator.    
	\item \emph{Controller-related attacks:} Attacks to high-level decision-making processes such as schedulers, dispatchers, and middleware services.
	\item \emph{Communication-related attacks:} Communication channels are principal targets of attackers. There are many ways to defend the channels,  primarily based on cryptography and coding. 
\end{enumerate}	
\subsection{SDCPS:``A High Level View''}\label{sec:overview}
\begin{figure}[h]
	\centering
	\includegraphics[width=0.45\textwidth]{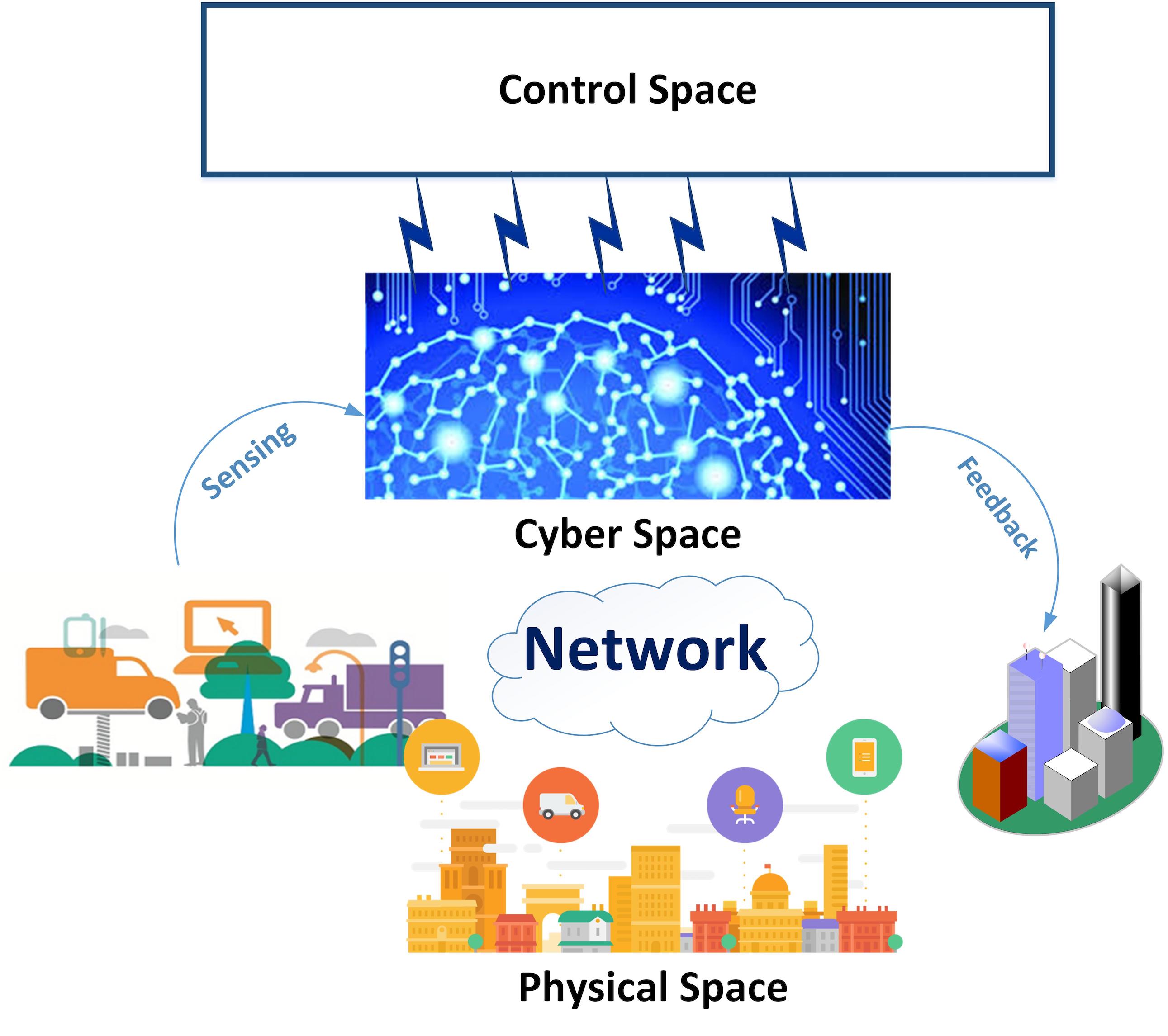}
	\caption{A `high-level' overview of the proposed solution.}
	\label{fig:BigPic}
\end{figure}

In this section, we present our SDCPS system architecture and illustrate how its  spaces and elements are structured within IoT systems. 
A general, `high-level' overview of the proposed software-defined model for CPS (SDCPS) is shown in Fig.~\ref{fig:BigPic}.  
The main aspects of the proposed architecture span three layers.

\begin{enumerate}[A.]
	\item \textbf{Physical space:} 
	
	Physical entities that need to be managed and controlled by IoT systems are enclosed in this space. Take as an example a \emph{smart home} 
	that comprises several devices like TVs, heating and AC, ovens, wash machines, doors, and many more entities that need to be controlled locally or remotely in an interconnected fashion. 
	Another example is in Intelligent Transportation Systems (ITS), where the physical space comprises cars, traffic lights, sensors, etc. The physical space is organized in domains and subdomains which can interact with the cyberspace via sensors, actuators, and dispatchers. 
	%
	
	\item \textbf{Cyber space:} This space encapsulates hardware and software designated for communication, sensing and information gathering and processing. It includes heterogeneous sensors, actuators and access-points. The quantity and quality of these devices depend on the IoT application under consideration. For instance, smart transportation systems may require more powerful processing units compared to smart homes. A formidable attribute of the abstraction in SDCPS is that no matter the type or number of devices or the application in place, the same control space can be installed and structured over the cyberspace to manage the physical space.

	\item \textbf{Control space:} This is the heart of the proposed architecture,  where all decision-making processes are initialized and taken. This space involves dispatchers, schedulers, security controllers and coordinators.

\end{enumerate}	

\begin{figure}[h]
	\centering
	\includegraphics[width=0.45\textwidth]{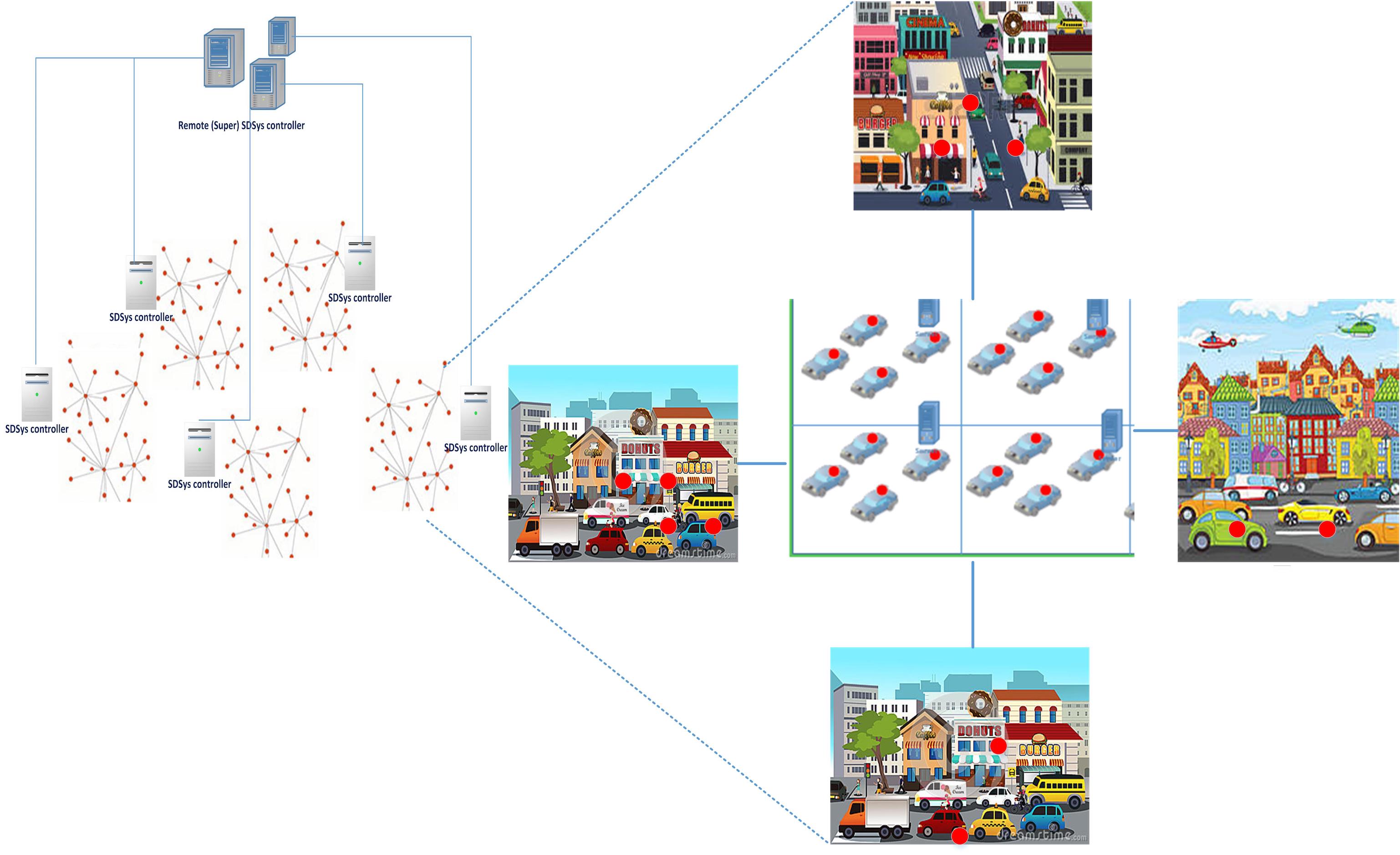}
	\caption{Distributed software-defined control layer.}
	\label{fig:fullyDis}
\end{figure}

The distributed control layer is the key point behind our proposed model, as  illustrated in Fig.~\ref{fig:fullyDis}.  The physical space is structured into several zones and sub-domains, where each one is controlled and managed by a corresponding  local controller. Local controllers communicate with one another to exchange information so as to carry out collaborative decisions. In transportation systems, for example, a given city can be divided into different areas, each one controlled by a local traffic control center, where the local controllers of different areas communicate information and control actions so as to improve experienced traffic  conditions throughout the metropolitan area.

\begin{figure}[h]
	\centering
	\includegraphics[width=0.45\textwidth, height=8cm]{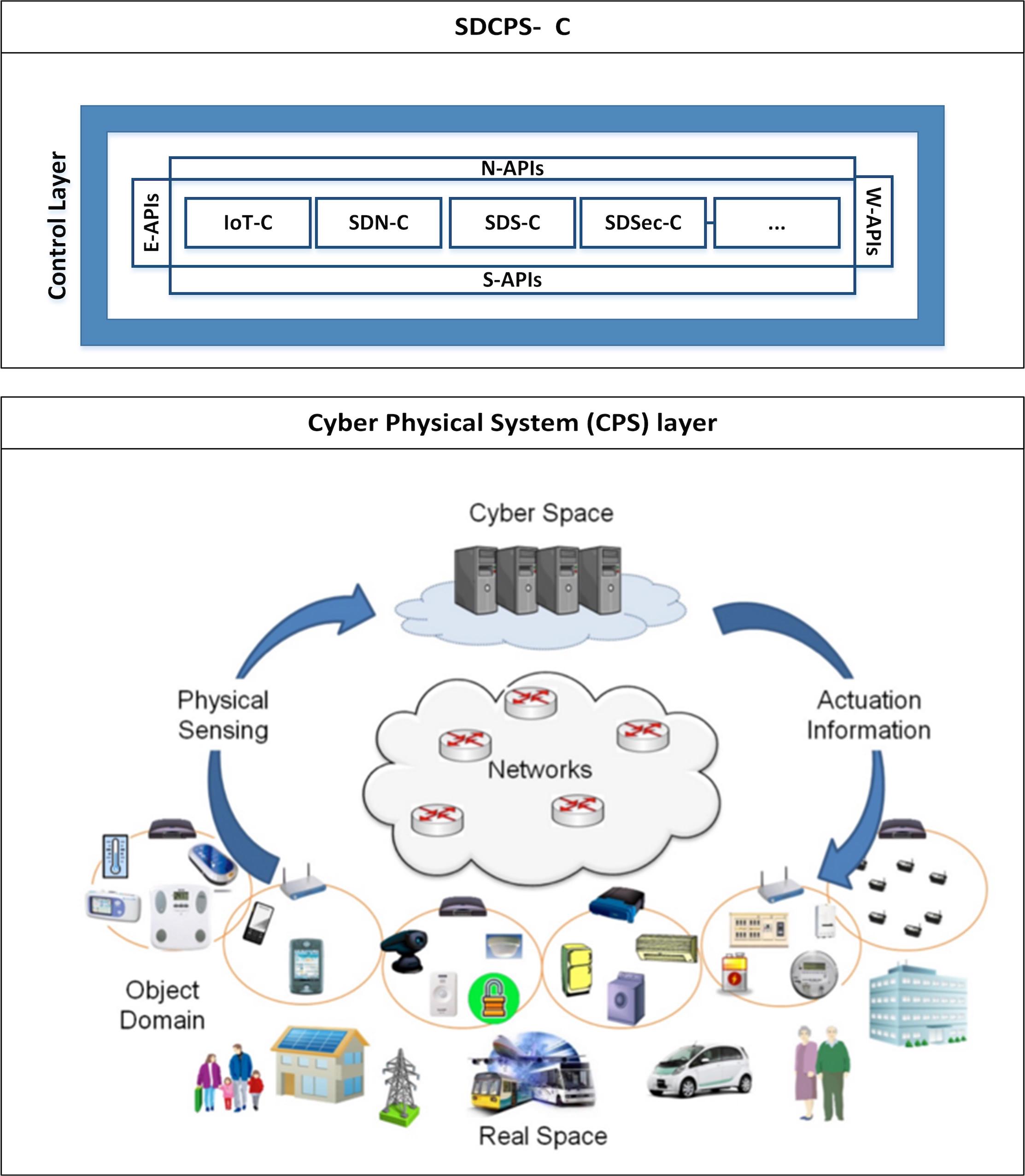}
	\caption{A partial sub-domain and its main components. The bottom figure is borrowed by the WiSE Laboratory at JAIST.}
	\label{fig:CPS_new}
\end{figure}

Fig.~\ref{fig:CPS_new} presents the building block for a multi-layer composition in which different sub-domains may be integrated into a vertical (i.e., hierarchical) or horizontal (i.e., decentralized) fashion.

\subsection{Architectural Elements}\label{sec:elements}

We present the main architectural components and elements of SDCPS along with their roles, responsibilities and interaction within IoT applications. To make an analogy with control systems, we use the language of linear\footnote{\label{note1}We use deterministic linear time-invariant systems with no loss in generality for simplicity of exposition. For the same reason, we assume an undirected communication graph.} dynamical systems:
\begin{subequations}
\begin{eqnarray}\label{linear}
\x^{(i)}(k+1) =& A_i \x^{(i)} (k) + B_i \u_i(k)\\
\y^{(i)}(k) =& C_i \x^{(i)} (k) + D_i \u_i(k),
\end{eqnarray}
\end{subequations}
where $\x^{(i)} (k), \u^{(i)} (k), \y^{(i)} (k)$ denote the state, measurement, and control input for plant $i$ at time $k$. We denote state-estimates (e.g., as obtained by Kalman filtering) by $\hat{\x}^{(i)}(k)$.
We further capture the communication network topology by an undirected\footnoteref{note1} graph $G(k) := (V(k),E(k))$ where two sub-systems $i,j$ may communicate at time $k$ if and only if $(i,j)\in E(k)$; the graph is time-varying, in general, due to agent mobility as well as the features of the wireless medium. For a node $i$, we define its neighborhood $\N_i(k) := \{ j: (i,j)\in E(k)\}$.


\para{Physical nodes}
\begin{enumerate}
	\item \textbf{Mobile node:} An entity with time-varying location due to mobility (take for example vehicles in a transportation network). 
	
	\item \textbf{Cluster:} The system can be partitioned into several clusters consisting of multiple atoms (mobile nodes). The set inclusion relation of each node to a cluster varies over time, as mobile nodes may partake in different clusters, primarily based on their location. 
		
\end{enumerate}

\para{Cyber nodes}
\begin{enumerate}
	\item  \textbf{Sensor:} A sensor gathers information from its surroundings 
	as prescribed by its sensing range. The resulting data is locally filtered and forwarded to the corresponding aggregate sensor. Each mobile node may possess several sensors. This corresponds to an entry (or subset of entries) of vector $\y$.
	
	\item \textbf{Aggregate sensor:} Each mobile node has one aggregate sensor that summarizes the sensing information gathered from on-board sensors. This is precisely what we have denoted as measurement vector $\y$.
	
	\item \textbf{Actuator:} 
	The information gathered is used by the system (through external upper control layers) and local controller to synthesize the control input of the actuator, i.e., $\u$. Decentralized actions amount to determining individual entries of $\u^{(i)}$ while distributed control laws describe strategies for synthesizing $\u^{(i)}$ from $\{\hat{\x}^{(j)}(k),\u^{(j)}(k)\}_{j\in \N_i}$. 
	
	
	\item \textbf{Access point:} Several sensors are linked to the system via wireless communication access points. 
	
\end{enumerate}

\para{Coordinators}
\begin{enumerate}
	\item \textbf{Local coordinator:} This controller is responsible to take actions for a subset of neighboring nodes by collecting data from the corresponding agents. 
	In the smart home example, we may organize the home into several local areas, i.e., rooms, each one managed by a local coordinator. In such case, several decisions like turning on/off the light in a room do not require any information from sensors in another room.  	 
	
	\item \textbf{Cluster Coordinator:}  Each cluster is assigned to a coordinator that applies roles and actions to the associated nodes within its sphere of influence. The main roles are shaped to manage and control nodes, transfer information from/to other cluster coordinators,  update the middleware (cf. Sec.~\ref{sec:middleware}) about the state of its associated nodes, and obtain and execute rules and instructions from upper layers through the middleware. As an example in the smart home application, several rooms are managed and controlled by a cluster controller to maintain balanced power supply among all of them.  	
	
	\item \textbf{Area coordinator:} In the smart home application this can be the floor coordinator, where several clusters are managed and controlled by this coordinator. This entity is a meta-controller which sets the rules for distributed and decentralized strategies, e.g., selecting from a subset of rules and privileges at each decision instant (i.e., how to design the feedback matrices $\{K_i$\} in our canonical example).

\end{enumerate}

\begin{figure}[h]
		\centering
		\includegraphics[width=0.45\textwidth]{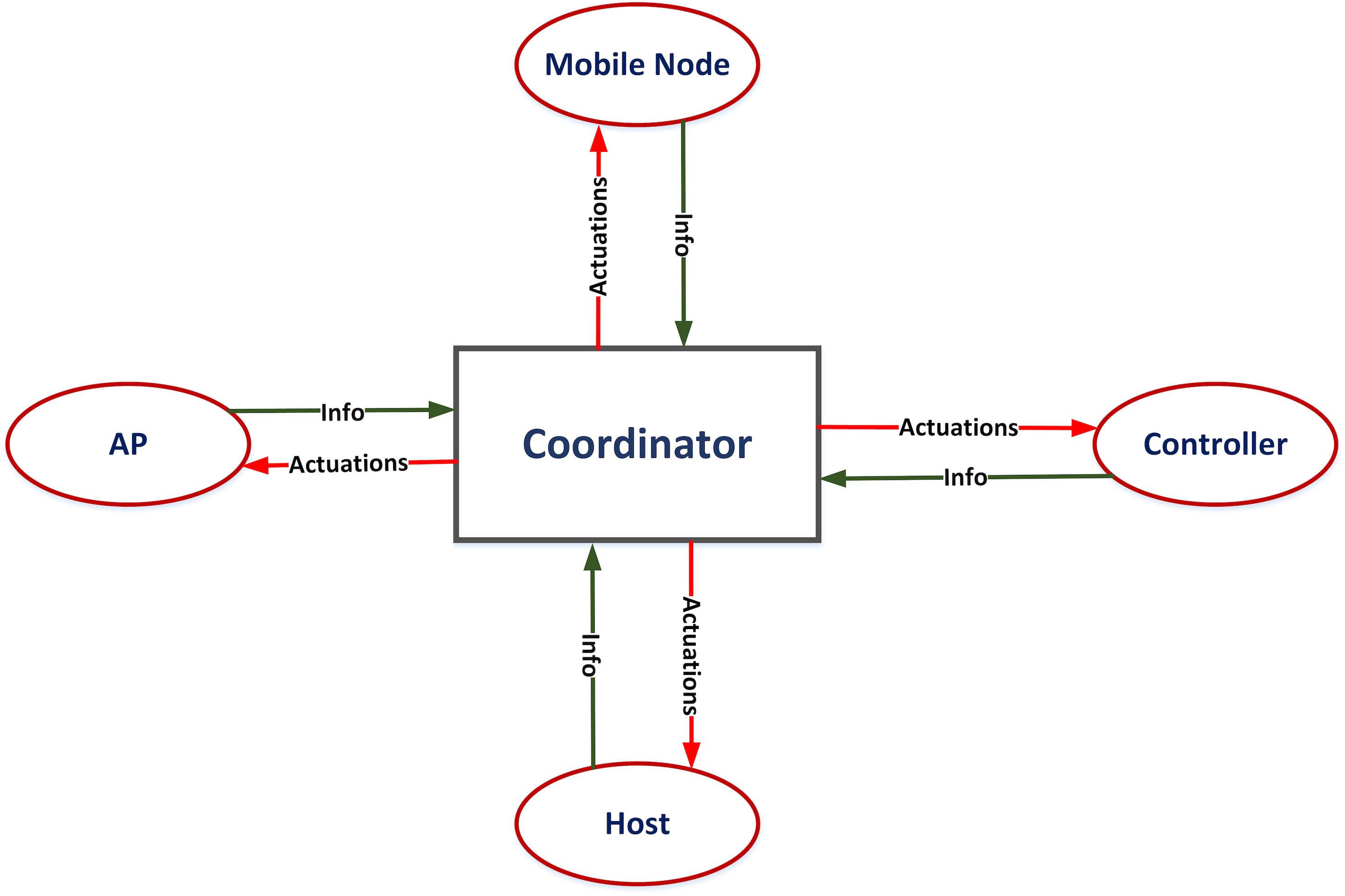}
		\caption{Communication Flow (Information and Actuation) through a coordinator; AP: access point.}
		\label{fig:Flows}
	\end{figure}

The coordinators can take actions based on acquired information from several resources as shown in Fig.~\ref{fig:Flows}.

\para{Controllers}
\begin{enumerate}
    \item \textbf{Self-Controller:} 
    Each mobile nod, such as a vehicle in the transportation system, has an autonomous controller which is called self-controller. It takes information from the aggregate sensors as input or, in some cases, may take higher commands from the local coordinator and controllers to launch a self-control decision.

    
    In our running example, self-controller may be an autonomous feedback control law, i.e.,
    $$\u^{(i)}(k) = K_i(k) \hat{\x}^{(i)}(k),$$
    such as in maintaining a constant velocity of a given vehicle, as determined by other coordinators and controllers.

    
    \begin{figure}[h]
    	\centering
    	\includegraphics[width=0.45\textwidth]{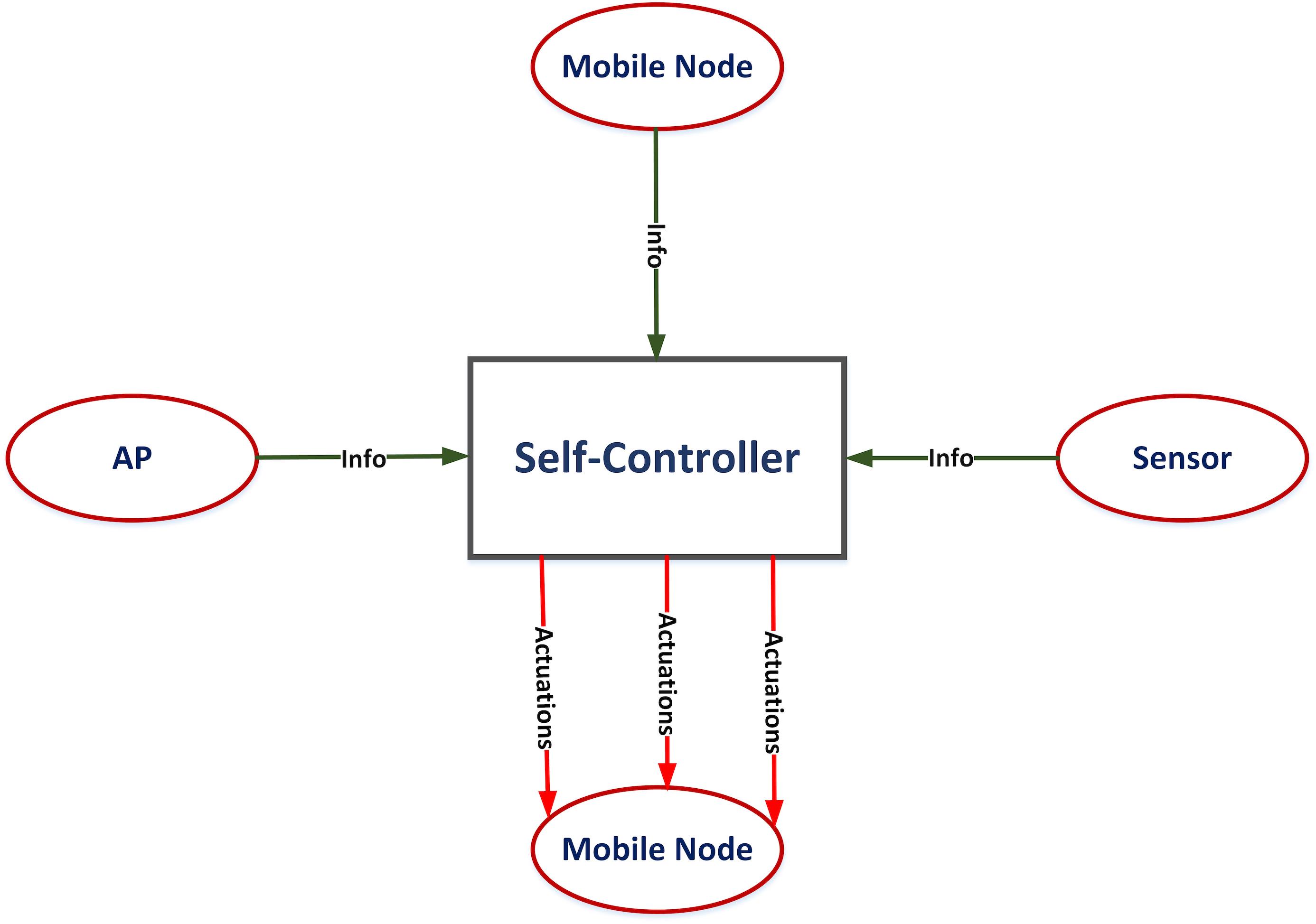}
    	\caption{Communication Flow (Information and Actuation) through self-controller; AP: access point.}
    	\label{fig:scontroller}
    \end{figure}

   	\item \textbf{Local Controller:} Local decisions that do not require permissions from upper-layer controllers are taken by the local controller. Information that is needed to take these decisions is acquired from aggregate sensors in the mobile node itself or from the ones in neighboring agents.

   	The distinction between the local coordinator and local controller lies in that the coordinator may make a decision related to more than one nodes (the coordinator defines control policies that a controller is responsible for enforcing).
   	
   	In control terminology, the local controller is a local feedback control of the form:
   	$$\u^{(i)}(k) = \sum_{j \in \N_i\cup i} K_j(k) \hat{\x}^{(j)}(k).$$
   	
   	\item \textbf{Super-controller:} It is to the local controller like the local controller is to the self-controller. It enables and supports the hierarchical decision-making process from top to bottom, while information flows from bottom to top.

   	In our example, the super-controller designs the feedback gains in a tree-like dependence.  
   	
   	\item \textbf{Global Controller:} This constitutes the engine in the proposed model, set in the root of the hierarchical model. All critical and high-level decisions are established at this controller. It has a holistic view of all remote nodes. 
   	    
\end{enumerate}

In the transportation system example, we may say that different parts of the car have their own controller like steering, engine and lights, where all such parts are controlled and managed by a local controller. The local controllers of all cars within a specific zone are controlled and managed by a super controller when there is a need to enforce a decision on that area, e.g., for collision avoidance purposes.

A set of design choices may be inferred to decide the number of layers and the depth of controller levels based on the specifics of the application under consideration. In this aspect, trade-offs are omnipresent: responsiveness vs. load balancing vs. security vs. complexity and so on. As an example, dividing the network into several clusters implies that the upper controller is responsible to control fewer lower level nodes. This may speed up the transfer of decisions and load balancing, but on the other hand it may increase the complexity of control synthesis and the vulnerability to a number of cyberattacks. Different scenarios may exist, no choices are absolute or clear in CPS, and everything has to be studied as part of a common whole.


\para{Interface: The Middleware}

Higher and lower coordinators and controllers are connected by a bridge, which is called the Middleware. It is the software residence for schedulers,  services, dispatchers and several software-defined controllers.

The main challenge in designing the middleware lies in maximizing scalability, adaptability and reliability. Our proposed architecture borrows principles from the  \emph{Etherware}~\cite{kim2008architecture}, destined for network control due to its real-time capabilities~\cite{kim2013real}.

The Middleware is composed of two types of \emph{components}: controllers and services. The former is a set of SD-controllers developed for control and coordination between the cyber and the physical, while the latter is responsible for managing the communication between the controllers and simplifying installation, development and execution. 
For real-time services, a \emph{real-time scheduler} is a key component of the middleware:  
%
packets and commands are assigned different priorities and execution deadlines that have to be accommodated by means of scheduling with QoS guarantees~\cite{hou2009theory}. 
A more detailed description of the middleware layer will be provided in Sec.~\ref{sec:middleware}.


\begin{figure}[h]
	\centering
	\includegraphics[width=0.45\textwidth]{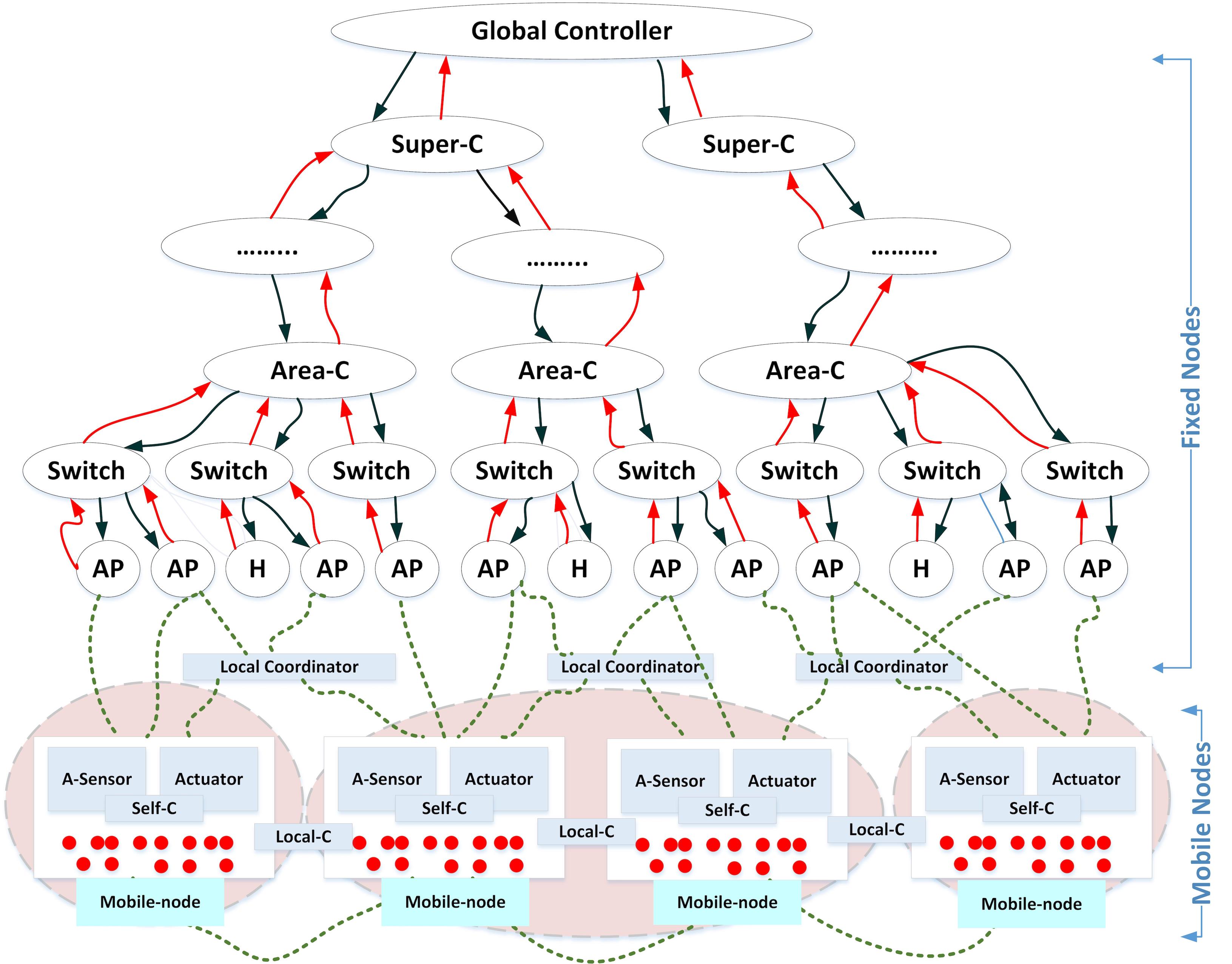}
	\caption{The components of software-defined control and the communication/actuation flow among them; H: host.}
	\label{fig:controlstr}
\end{figure}

Fig.~\ref{fig:controlstr} abbreviates the interactions and interconnections between system components.

	\subsection{Control architecture}\label{sec:control}

\begin{figure}[H]
		\centering
		\includegraphics[width=0.45\textwidth]{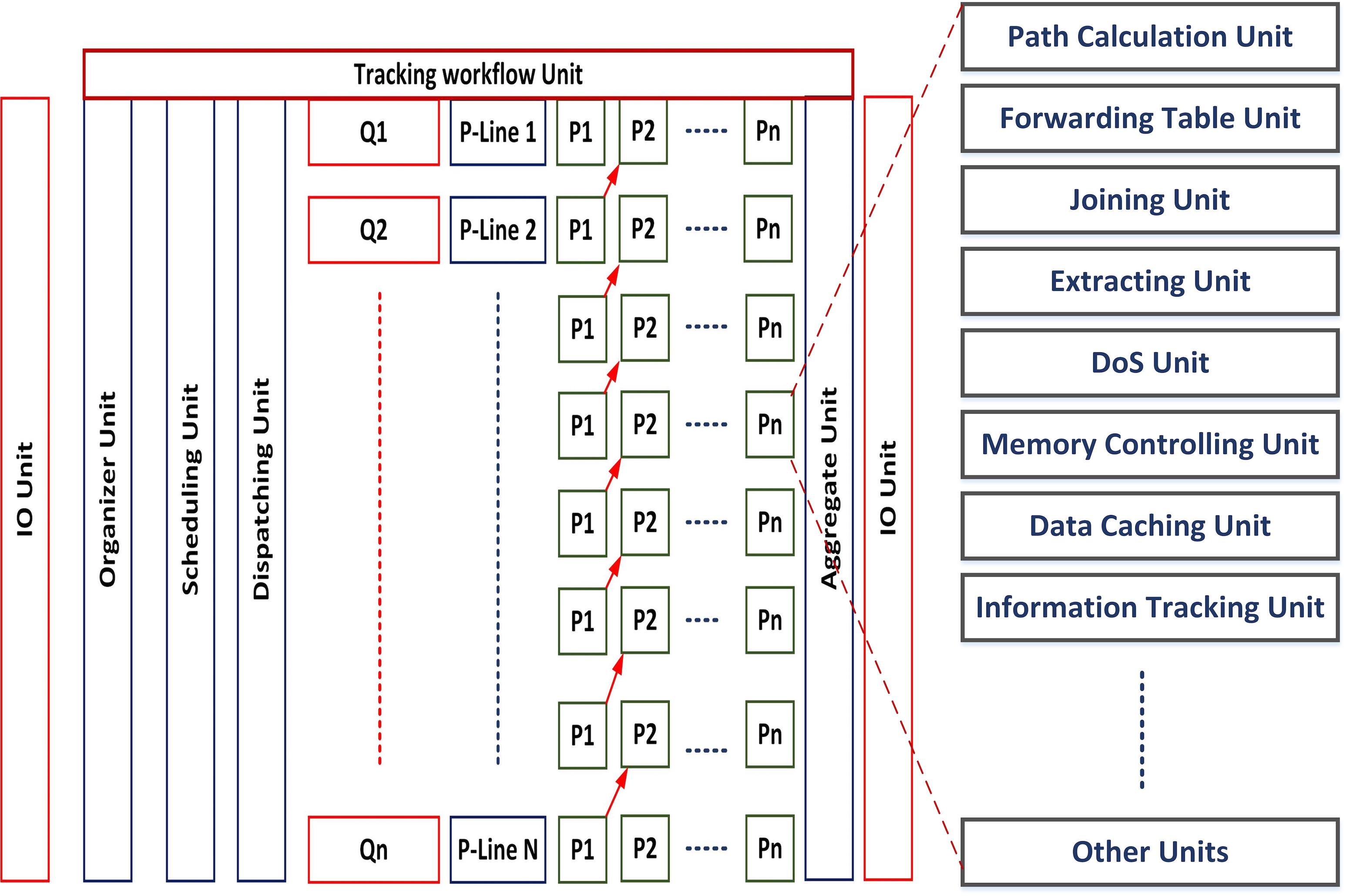}
		\caption{The control architecture and its main units.}
		\label{fig:ContUni}
	\end{figure}

In this section, we zoom in to discuss in further the units and components of the proposed model; see Fig.~\ref{fig:ContUni}.

\para{Physical structure} 

The physical structure of each controller consists of several processors, and each possessor is mapped to multiple processing units. The controller can run several processes simultaneously in a multi-threading fashion. Moreover, CPUs and GPUs can be leveraged, and one-pass controller is used, as it is typically faster and simpler than its multi-pass counterpart, and it helps improve reliability, security and system performance. Pipes between producer/consumer threads are used as safeguard communication channels for security purposes.

\para{Logical structure}

\begin{figure}[h]
	\centering
	\includegraphics[width=0.45\textwidth]{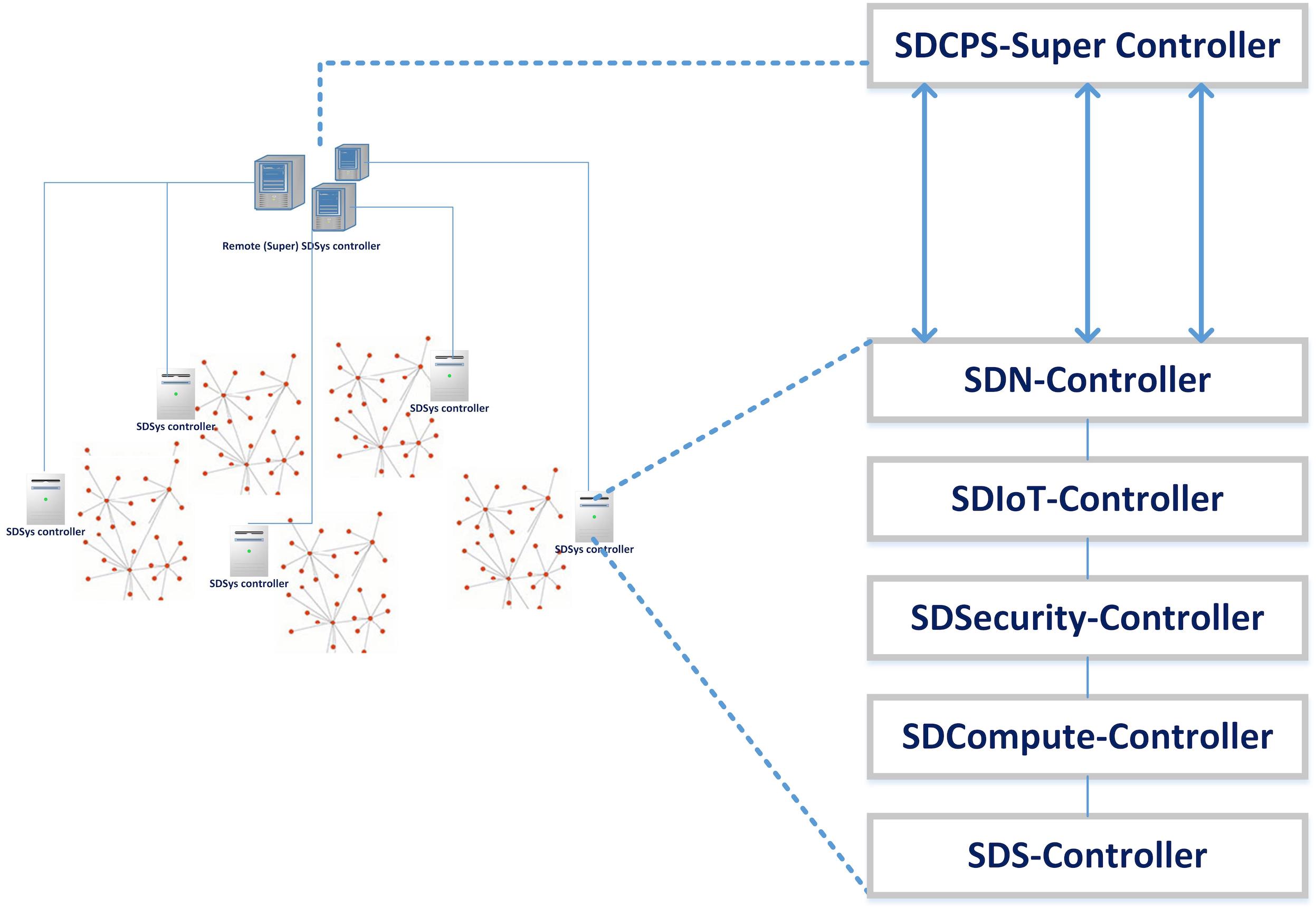}
	\caption{The main controllers inside the SDCPS\_Controller.}
	\label{fig:ConUnits}
\end{figure}

 A set of dedicated software-defined (sub-)controllers are installed in each controller, which communicate and collaborate with one other to maintain a smooth work-flow. 
 Fig.~\ref{fig:ConUnits} illustrates the logical view. 
 
 Each one comes equipped with several units which we present and define in the following. Precisely, we show how these units alongside their main responsibilities  work together to control and manage IoT applications in a software-defined manner.

	\para{SDN\_Controller:}
	This controller is responsible to control and manage the networking part of the system. It maintains network information for all nodes within its range, which is used for forwarding-table generation. These tables are forwarded to the switches that are connected to this controller, for multi-hop communication purposes. Different algorithms can be followed to generate these tables like minimum spanning tree, shortest-path, etc.), depending on the target of each IoT application. For instance, in transportation systems, real-time decisions have to be taken and for such situation, algorithms that specifically target real-time scheduling will be advantageous.

	The main units inside SDN Controller are:
	\begin{enumerate}
		\item \textbf{Path calculation unit:} Calculating the path from source to destination is the responsibility of this unit. Such a path is taken based on several criteria. 
		
		\item \textbf{Forwarding table generation unit:} All forwarding tables are generated inside this unit based on the path calculation unit outcome. These tables are generated and stored by this unit.  It also gets feedback from the network status tracking unit to adjust the calculated path when needed,  as explained next.

		\item \textbf{Network status tracking unit:} Any change (bottlenecks, channel degradation or broken links) in the status of the network is tracked by this unit. For instance, in a transportation network, when a new road is built in a specific region this unit receives a notification about this update to take further actions.

	\end{enumerate}
	
	Other units for enabling the joining and extraction of nodes are also available.
	\\ 
	\para{SDIoT\_Controller:}
	
	All information about IoT application smart devices is kept here. Information such as logical status and physical location is used in the coordination and management process.

	\begin{enumerate}
		
		\item \textbf{Smart device information tracking unit:} The role of this unit is to scan and track the smart devices for requests that have to be communicated to the various controllers.
		
		\item \textbf{Smart device status tracking unit:} This is responsible to track the status of a smart device as relevant to communication and actuation (for example, busy, sending, receiving, `asleep' in duty-cycling, low battery, etc.).		
		
		\item \textbf{Smart device location tracking unit:} As the mobile nodes are moving, the physical location of smart devices also varies. It is crucial to track the location continually and accurately to enable location-aware services and controls (such as set inclusion in area coordinators, and distributed communication, computation and control). 
	\end{enumerate}
	Smart device joining and extracting processes are also the responsibility of this controller.

	\para{SDSecurity\_Controller:}
	
	The responsibility of keeping a secure system is assigned to this controller. It has a set of units with specific mechanisms and tools that are interconnected to each other to detect, prevent and resolve several types of cyberattacks.   
	
	\begin{enumerate}
		
		\item \textbf{Auditing and Mapping unit:} It is important to keep information about the switches, routers, access points and other network nodes for security and safety purposes. This unit takes the responsibility for auditing all relevant information about the network infrastructure nodes like its vendor, location, and type, in order to discover all abnormal behaviors.

		\item \textbf{Knowledge-based unit:} This unit is responsible to store and keep all discovered attacks in the system. In case a new attack is discovered, the inline mechanisms and tools are used to determine the required solution, while this attack is stored in the unit.

		\item \textbf{Scanning and Screening unit:} It uses several mechanisms and tools to scan traffic and detect if there is a threat to trigger the resolution unit.  Many scanning tools are available, each one responsible to scan a specific type of information.

		\item \textbf{Monitoring unit:} The tools in this unit help network defenders discover and analyze anomaly activities in the network. 
		For this purpose, there are many visualization tools, and an unceasing need to develop real-time monitoring methods. 
		
		\item \textbf{Detection unit:} This unit works with other controller units to detect any type of cyberattacks. 
		There are a lot of existing tools for this purpose: MINDS, ADAM, NIDS being notable examples.

		\item \textbf{Prevention  unit:} This tool is responsible to prevent the attacks from inducing any anomalous actions and spreading them over other CPS domains.

		\item \textbf{Handling unit:} The tools of this unit are responsible to resolve and handle the attacks in case that they cannot be prevented, or when the detection was too late. It tries to eliminate and resolve its effects and then notify the knowledge-based unit for this type of attack.

		\item \textbf{Policy unit:} A set of security policies correlated to CPS are sustained inside this unit. Dedicated policies for each IoT application are defined and applied.  For example, in a smart home, the user can set policies to keep the room temperature no more than a given threshold and increase it only in specific situations. 
		
		\item \textbf{Security status tracking unit:} 
		This unit recaps the status of the system as related to security. 
		
		\item \textbf{Encryption/Decryption unit:} Encrypting traffic is considered an efficient way to protect packet-based communication from intruders. This unit is responsible for ascertaining data privacy by means of applying encryption/decryption methods on the data. 
		Nonetheless, we note that encryption introduces storage and time overheads (transmission and processing) and cannot be used as a passepartout, in particular for time-critical applications.
		
		\item \textbf{Backup unit:} Keeping backup versions of the system status facilitates its effective protection. This unit keeps frequent back-ups 
		to restore the system when the effects of an attack cannot be resolved otherwise. 
		
		\item \textbf{Up-To-Date unit:}  It has a set of procedures to update the existing software to tackle new attacks with new solutions.

	\end{enumerate}
	\para{SDCompute\_Controller:}
	
	Network computation resources like CPUs, GPUs, and RAMs are controlled and managed by this controller. Dedicated tools for GPU, CPU and memory managements are installed.

	\para{SDS\_Controller:}
	This controller manages storage devices and processes.  
	
	\begin{enumerate}
		\item \textbf{Data storing unit:} This unit takes the responsibility of controlling the data storing process in the storage arrays. 
		\item \textbf{Data caching unit:} In large-scale CPS where the hosts are distributed over a large area, it is important to cache parts of the most frequently requested/used information locally~\cite{SDCache}.
    
		\item \textbf{Data de-duplication unit:} In distributed database systems, it is important to maintain concurrency of information; this unit accounts for eliminating duplicate values so that any user gets access to the most up-to-date available information.
		
	\end{enumerate}

\para{Shared units by all types:} 
Some commonplace units that are available by almost any type of controller include: 

\begin{enumerate}
	\item \textbf{I/O Unit:} The input/output unit 
	sends, receives and forwards packets from and to other units. 
	\item \textbf{Organizer Unit :} This unit assigns priorities to packets based upon predetermined QoS criteria. 
	\item \textbf{Scheduling Unit:} This unit schedules packet transmission taking into account the assigned priority; there is a great number of scheduling algorithms that can be used for this purpose, see for example~\cite{hou2009theory}. 
	\item \textbf{Aggregate Unit:} This unit is responsible for aggregating  packets of a given flow before forwarding them to the next processing unit. 	
\end{enumerate}

	\para{SDCPS\_Controller:}
	This controller comprises the main engine of the system. It is responsible to effectuate all of the functionality described in the prequel and organize the interplay of the phyical and cyber space by overlooking and coordinating all other controllers in the system, either directly or indirectly.
	

\subsection{Middleware Layer Architecture}\label{sec:middleware}

\begin{figure}[H]
	\centering
	\includegraphics[width=0.45\textwidth]{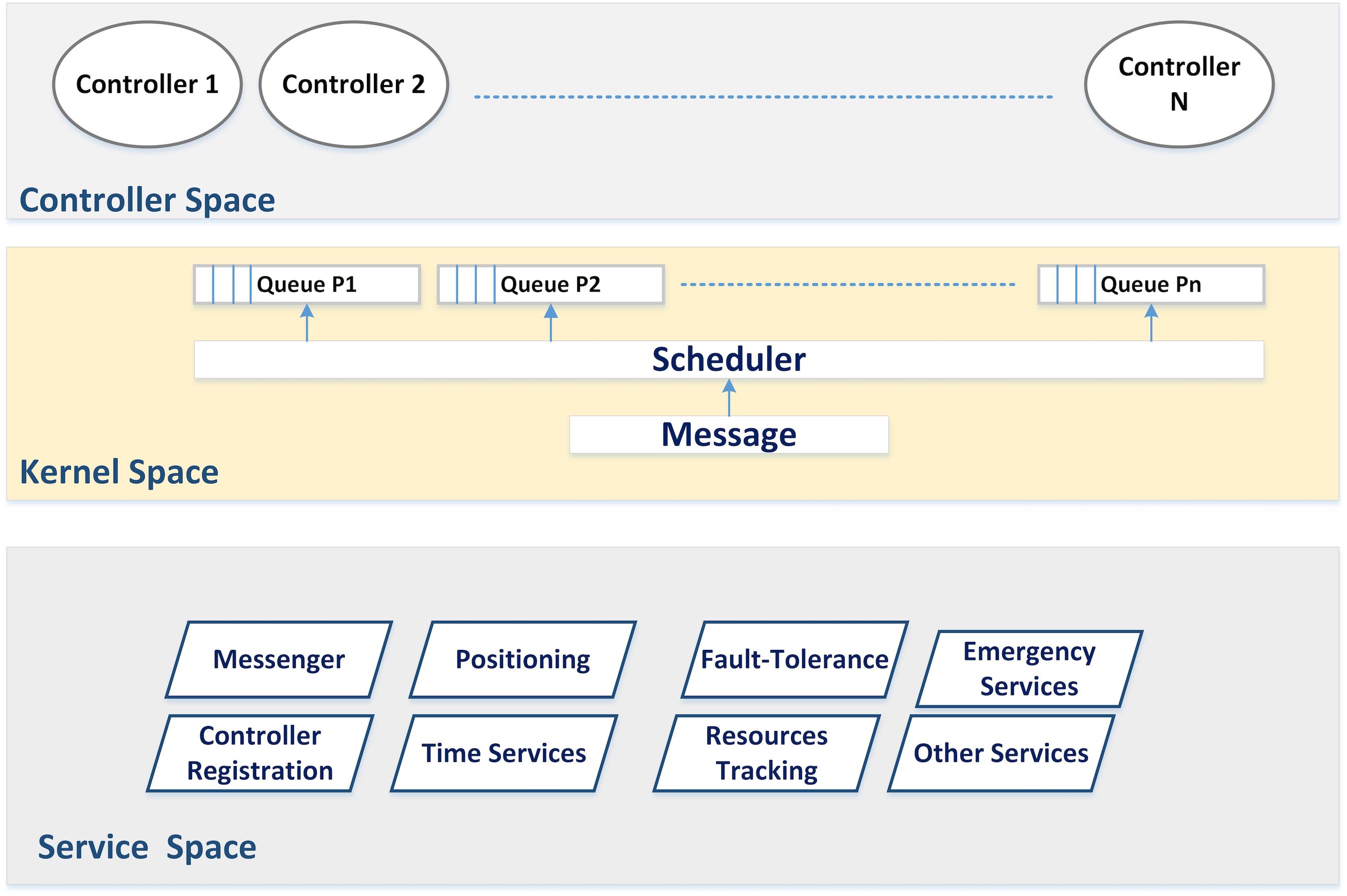}
	\caption{The Middleware layer and its main components.}
	\label{fig:middleware}
\end{figure}

The middleware layer is configured to accelerate real-time decision making and facilitate the communication and interactions between system control layers.     
All services and entities are software-defined, which empowers the modifications and component migration processes on-the-fly. Fig.~\ref{fig:middleware} shows the structure of the middleware layer and its three spaces: controller space, kernel space, and services space.

\para{Controller Space:} Each controller is implemented as a component itself, and all components interact with each other as explicated before. \\

\para{Kernel Space:}  This space is responsible to schedule packets based on their priority and time-criticality, as assigned by the tools of the real-time controller.  For instance, a signal coming from an ambulance or firetruck should be served first.    
A wealth of scheduling algorithms ~\cite{hou2009theory} can be implemented. Each IoT application has different QoS requirements, and based on these requirements the appropriate scheduling algorithm will be selected.

 Post scheduling decisions, a packet is assigned to a specific queue based on its priority, but also the physical positions of the mobile sender/receiver. It is therefore essential to provide position tracking service for location-aware real-time scheduling. Scheduling queues in a network is well-studied~\cite{srikant2013communication}, and extensions are possible for optimal scheduling subject to deadlines and priorities~\cite{hou2009theory}.

 \para{Service Space:} Various services are encapsulated inside this space to enable real-time decision making and improve network communication:

 \begin{enumerate}
 	\item \textbf{Messenger Service:} This service keeps a smooth communication between all controllers at the same layer through the west/east APIs in Fig.~\ref{fig:CPS_new}; take for example controllers within the same floor in a smart building. Additionally, it handles the communication between control layers located at different floors through SDN.

 	\item \textbf{Position Tracking Service:} The positions of mobile nodes over time are tracked by this service. Note that node position may be changed as a result of a control decision (e.g., in a transportation system a local controller routes a car at an intersection).

 	\item \textbf{Speed Stamp Service:} Estimating and computing mobile nodes' speed is useful to predict its position over time (e.g., by using tracking tools such Kalman filtering), for example to predict when a can car will reach a specific route to avoid collision or deadlock.

 	\item \textbf{Fault-Tolerance Service:}  For large-scale IoT systems fault-tolerance is crucial. A set of solutions and policies are implemented inside this service to recover from hardware/software failures (i.e., how to switch to a nearby controller in case the local controller fails).   
 	
 	\item \textbf{Controller Registration Service:} This service is responsible to insert a new controller along with its relevant information so as to keep the  system updated.

 	\item \textbf{Time-stamping service:} A dedicated unit is implemented inside the middleware layer to record the times of past, present and future events such as sensed information and control actions.

 	\item \textbf{Time-translation service:} In a large-scale CPS, clocks don't agree~\cite{freris2011fundamental} and it is fundamental to convert packet' time to controller's time, and vice versa.

 	\item \textbf{Time synchronization service:} Accurate clock synchronization is a instrumental for distributed coordination in CPS. It affects both performance as well as safety~\cite{freris2011fundamental}, with examples enumerating wireless protocols such as MAC, duty-cycling, formation control, concurrency in databases, and more. To this end, algorithms with high scalability, and low communication and computation overhead are especially important to implement in the middleware~\cite{RK,RK2,nfrer_algocdc}.
 	
 	\item \textbf{Resource Tracking Service:} This service targets tracking the resources in the entire system with the intention of providing load balancing and fairness among users.

 	\item \textbf{Emergency services:} This set of services tracks components and takes actions when they become unavailable or fail.
 	
 \end{enumerate}


	
\subsection{Communication \& Control Planes}\label{sec:flow}

\begin{figure}[h]
	\centering
	\includegraphics[width=0.45\textwidth]{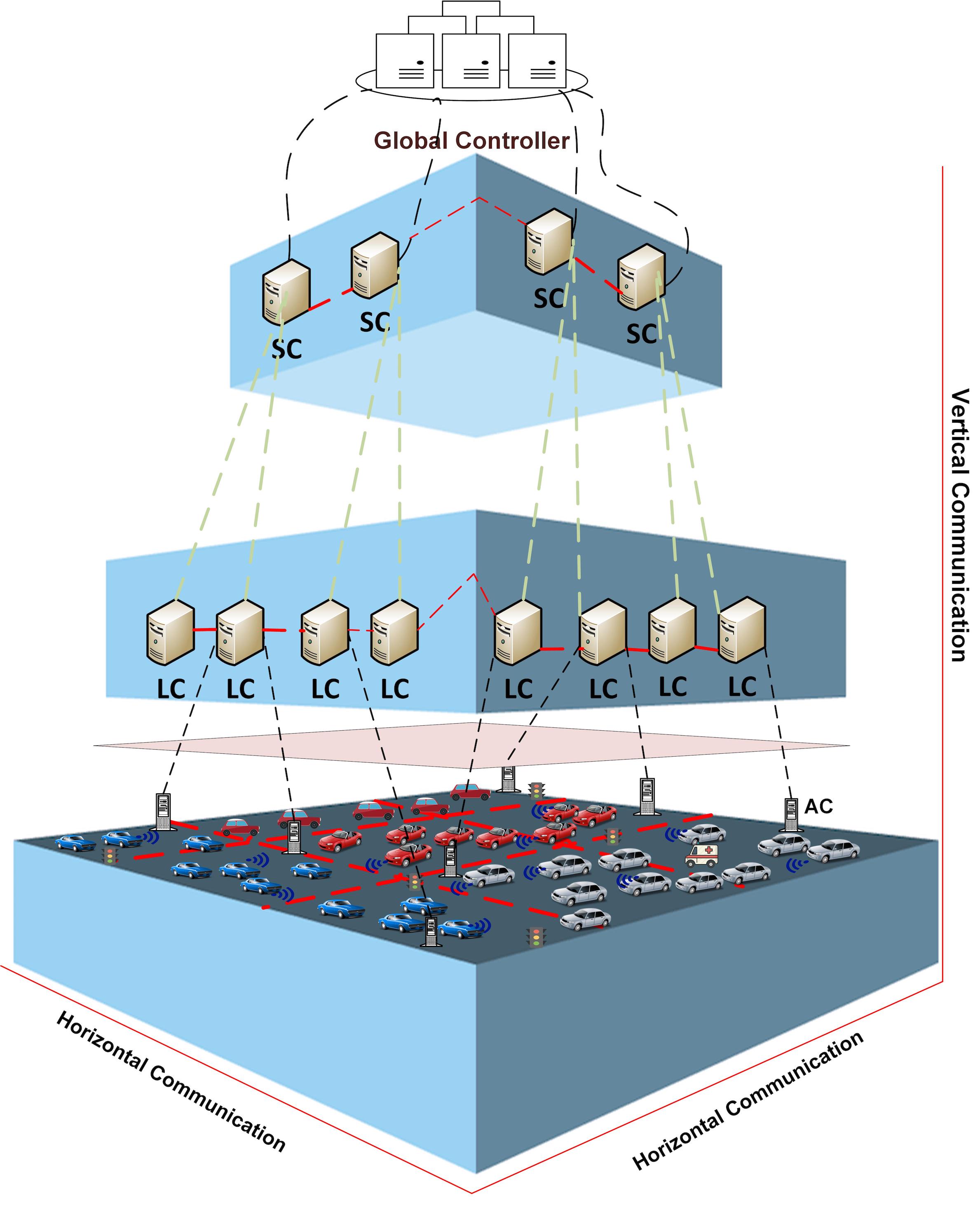}
	\caption{The Horizontal and the Vertical control planes.}
	\label{fig:CPSlayers}
\end{figure}

The proposed architecture features an interplay between two planes of communication and actuation, namely \emph{horizontal} and \emph{vertical}. Fig.~\ref{fig:CPSlayers} illustrates a global view of these planes, further elaborated in the sequel (note that we use the transportation application for demonstration purposes, which explains the cars in Fig.~\ref{fig:CPSlayers}).

\begin{enumerate}
	\item \textbf{Vertical Plane (Control Flow):} As explicated above, control flows in the network in a hierarchical fashion. 
	%
	At the very top level resides the root controller, while all mobile nodes (such as vehicles in a transportation system) reside at the low level. 
	The upper layer has coordination power over its immediate lower layer, and higher  privileges in resolving conflicts. 
	A middleware layer is responsible for combining and coordinating the controllers at the same layer and managing the communication across different layers. The number of layers at the vertical plane is closely related to  
	the size of the network at the horizontal plane  (e.g., it depends on the sizes of the area clusters such as area dispatchers in urban transportation).            
	\item \textbf{Horizontal Plane (Data Flow):} This plane reflects the communication between mobile agents and controllers within the same layer (for example direct communication among nearby vehicles, and among dispatchers at the same layer). 
	Recall the trade-off between the partition sizes and vertical plane depth: when nodes are horizontally clustered to smaller groups (areas), a larger number of cluster coordinators and area coordinators is required; this affects the accuracy, responsiveness, performance and security of the entire system, as well as infrastructure costs, that should all be considered by system designers.
\end{enumerate}

\para{Decision Making Process}\label{sec:decision}

Decision making process is performed in a systematic way by the proposed system. Such rethinking promotes the real-time decision making process, system reliability, security and scalability.  A brief discussion on control packet flow is explained in the following.  

\subsubsection{Centralized vs. Decentralized / Distributed}

An effective control and management plane is a key to system proliferation. Distributed control is integral for handling huge network traffic volumes via decentralized autonomy that achieves a scalable, sustainable, robust and adaptable system operation in a highly dynamic environment (take for example the failure of a traffic light or closure of a route due to an accident or natural disaster). However,  ``to distribute or not to distribute?'' does not always have an easy answer: local communication and interactions may cause a conflict through several system levels,  which require a supervisory control to solve it, whose complexity increases as the  network size is growing.  The proposed architecture aims to leverage the benefits of both worlds (centralized and decentralized / distributed) in a coherent methodological fashion.

\subsubsection{Flow of control packets}

The proposed methodological way of the decision making process fundamentally influences the control packet flow. Information is harvested from underlying control layers and is communicated to controllers in higher layers that take higher-level  control decisions and actions. Simultaneously, controllers in the underlying layers need to collaborate and communicate with each other and they are responsible to control all system entities assigned to them by taking the proper actions and forward the information to super controllers at a higher layer. All and all, this is treated diversely in different contexts: for instance, the auditing and screening security unit within the SDSecurity controller, previously described, is installed and implemented on all software-defined controllers, but does not perform the exact same operations in each component. To conclude, decision-making can be categorized ( among others
) to: 

\begin{itemize}
	\item \emph{Self decisions}: A given node can take a decision by itself without consolidating with other controllers or coordinators at the same or higher levels. 
	
	\item \emph{Coordinated decisions}: Nodes are not fully autonomous in taking a certain decision locally, and coordination with higher layers is needed to take further actions.

\end{itemize}

\begin{figure*}[!ht]
		\centering
		\includegraphics[width=0.96\textwidth]{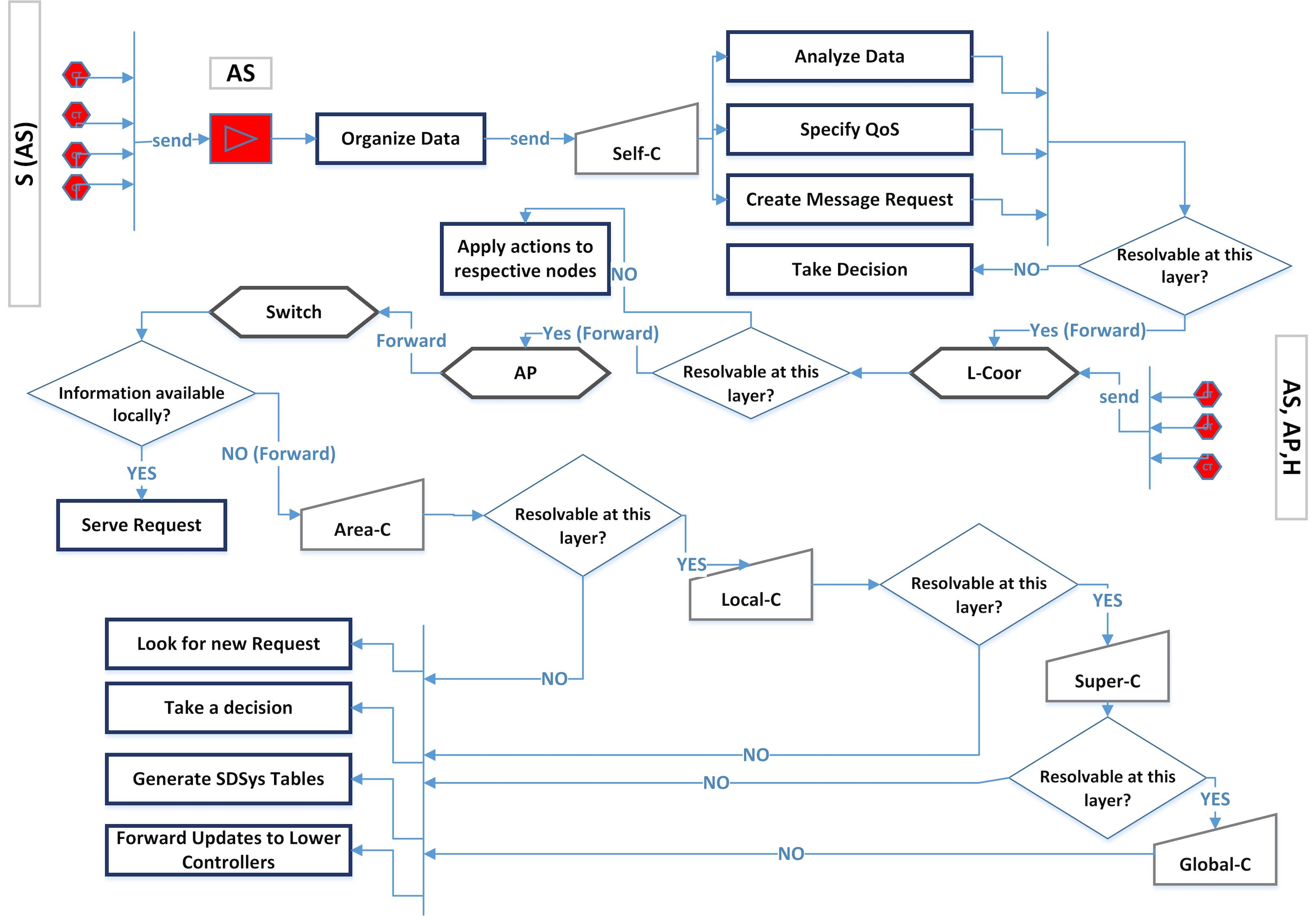}
		\label{fig:workflow}
		
		\caption{The workflow diagram of processing control requests in the SDCPS framework.}
		\label{fig:active1}
	\end{figure*}

This taxonomy applies inductively to all levels of the decision-making hierarchy; cf. Fig.~\ref{fig:active1} illustrates the work-flow in the system. 
	
%
%
		
\begin{figure}[H]
			\centering
			\includegraphics[width=0.4\textwidth]{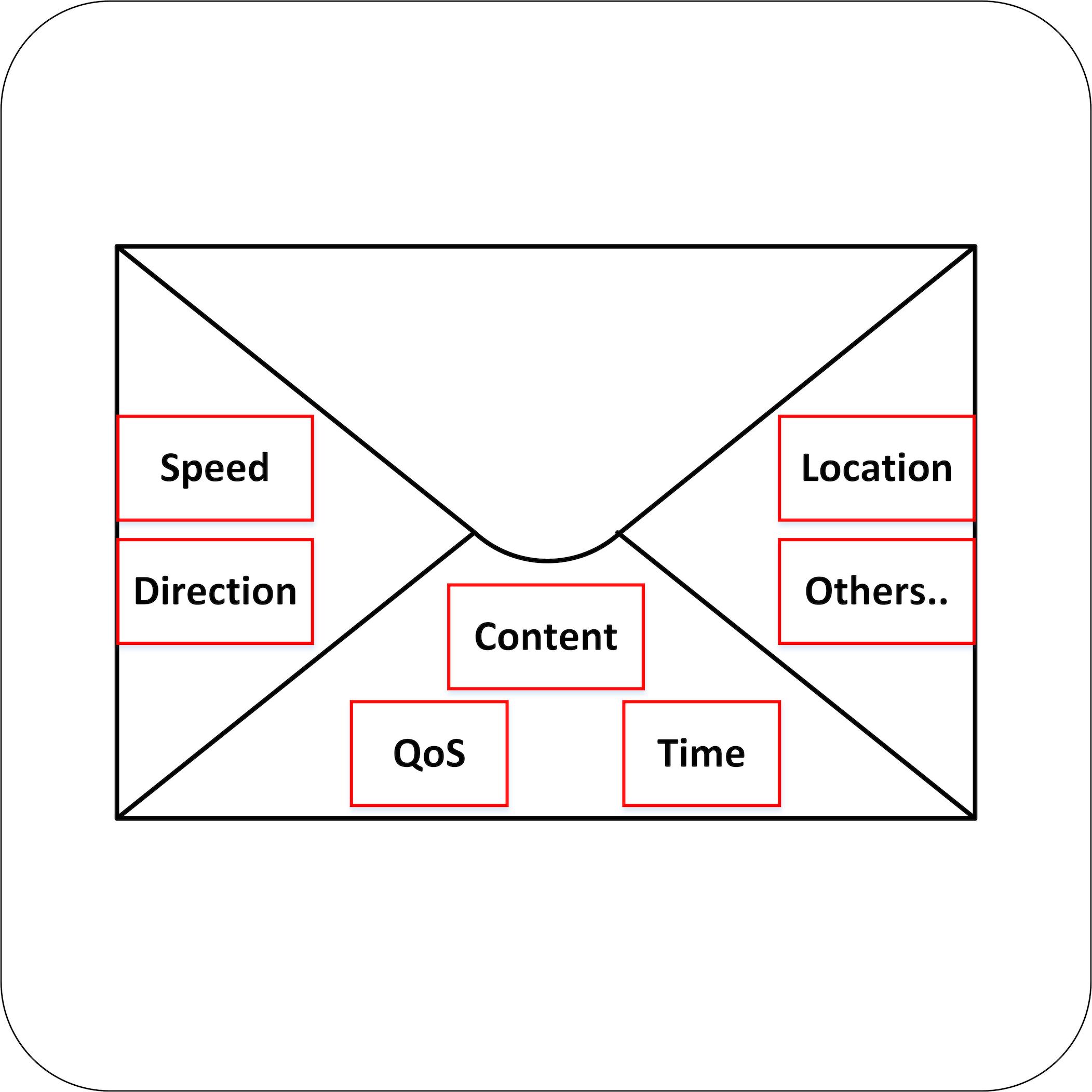}
			\caption{The contents of a packet in SDCPS.}
			\label{fig:message}
		\end{figure}

		\para{Packet Structure:}\label{sec:packet}
		To support the proposed architecture, it is also important to define the packet attributes required for control, communication and scheduling; for example agent-specific positioning and timing information, packet priority and many others; cf. Fig.~\ref{fig:message}.
		
%
		
	The setup of the system is exposed in Algorithm~\ref{Algo1}.


	\begin{algorithm}[h]
		\caption{The setup algorithm for the proposed model}
		\label{Algo1}
		\begin{algorithmic}[1]
			\Procedure{Setup}{}
			\State \textbf{Start}
			\State \textbf{STEP1:} Run and initialize Global Controller.
			\State \textbf{ControllerLIST[]= None}
			\State \textbf{SuperConLIST[]= None}
			\State \textbf{LocalConLIST[]= None}
			\State \textbf{AreaConLIST[]= None}

			\State \textbf{STEP2:} Create controller objects and initialize them. 
			\State \textbf{ControllerLIST[]} $\gets$ Create \& Initialize list of  controllers.
			\State \textbf{Create instants from the SD Controller and assign an ID for each one}
			\State \textbf{for $i=0$, $i{+}{+}$,while $i < length(instants)$}		
			\State \textbf{\qquad Controller LIST[i]} $\gets$ instant(i)\_ID
			
			\State \textbf{STEP3:} Assign types for each controller and create subsets.		
			\State \textbf{SuperConLIST[] $\subseteq$ ControllerLIST[]}
			\State \textbf{LocalConLIST[] $\subseteq$ ControllerLIST[]}
			\State \textbf{AreaConLIST[] $\subseteq$ ControllerLIST[]}

			\State \textbf{STEP4:} Global Controller partitions the system environment into sub-domains and assign all the controllers. 
			\State \textbf{PartitionLIST[]} $\gets$ Create \& Initialize list of Partitions.		
			\State \textbf{Assign area controller for each partition}
			\State \textbf{for $i=0$, $i{+}{+}$,while $i < length(ContollerLIST)$}		
			\State \textbf{\qquad PartitionLIST[i]} $\gets$ AreaConLIST[i].
			\State \textbf{}
			\State \textbf{Assign \textit{``Area Controllers"} to associated Local controllers}
			\State \textbf{Assign \textit{``Local Controllers"}  to associated Super controllers}
			\State \textbf{Assign \textit{``Super Controllers"} to Global controller}

			\State \textbf{STEP5:} Configure all software-defined units inside each controller.
			\State \textbf{for $i=0$, $i{+}{+}$,while $i < length(ContollerLIST$)} 
			\State \textbf{\qquad \While{$ ContollerLIST[i].units \not=Established$}}
			
			\State \textbf{\hspace{30pt} SDN\_Unit} $\gets$	Forwarding\_Table
			\State \textbf{\hspace{30pt} SDIoT\_Unit} $\gets$ Sensors\_Table
			\State \textbf{\hspace{30pt} SDSecurity\_Unit} $\gets$ Policies\_Table
			\State \textbf{\hspace{30pt} SDStorage\_Unit} $\gets$ Storage\_Table
			\State \textbf{\hspace{30pt} SDC\_Unit} $\gets$ Compute\_Table 
			
			   \State \textbf{end while}
			\State \textbf{STEP6:} Run the Controllers. 
			
			\State \textbf{STEP7:} Root Controller takes images from all controllers and keep the state for each one. 
			\State \textbf{for $i=0$, $i{+}{+}$,while $i < length(ContollerLIST$)}  
			\State \textbf{\hspace{30pt}  Image} $\gets$ ContollerLIST[i].captureState
			\State \textbf{\hspace{30pt} Root.Receive()} $\gets$ Image
			\State \textbf{STEP8:} Root Controller forwards all controller images to underlying controllers.        
			\State \textbf{End}
			\EndProcedure
		\end{algorithmic}
	\end{algorithm}

	\subsection{The features of the proposed model}\label{sec:features}
	The main features of the proposed architecture are briefly summarized in what follows: 
	
	\begin{enumerate}
		
		\item \textbf{Real-time:} The proposed system architecture promotes real-time decision making in three ways: (a) adding a middleware layer to facilitate and scale the communication and interaction between various system layers, (b) implementing services and controllers as self-components at the middleware, (c) applying QoS-based packet scheduling over multiple levels.      
		
		\item \textbf{Reliability and Security:} 
		Service availability, in the presence of malicious users, can be guaranteed by prescribing ways of taking control actions and accessing data. This can, in turn, be achieved by sharing and distributing responsibilities over several system modules, and employing redundancies that improve resilience to single points of failure or attack.    
		
		\item \textbf{Flexibility and Scalability:} 
		In volatile large-scale CPS, high-frequency control and management is required as some nodes may undergo failure or cyber attack. Such considerations can be served by our proposed solution in a fast, simple and self-configured programmable way. Moreover, our design combines decentralized distributed sensing, information retrieval, and control to guarantee scalability even for millions of mobile nodes.

	\end{enumerate}

\section{Experimental results}
\label{sec:exp}

In this section, we describe our emulation testbed 
and expose several experimental findings to underline the merits of adopting software-defined control procedures for IoT applications, in terms of efficiency and adaptability.  

\subsection{Emulation Environment}
Our testbed environment was implemented by installing Mininet 2.2.0rc1 VM~\cite{Mininet} over the Oracle Virtual Box and remotely linking it with the Ubuntu OS. We used Python as the programming language 
and extended the POX controller of Mininet (namely User\_Switch) along with its host main classes so as to capture several elements of the proposed architecture. 

\subsection{Test Scenarios and Experiments}

We have tested a range of scenarios for variable network sizes and topologies.   
The network topology is captured by a graph $G=(V,E)$ consisting of $n$ vertices (controllers and end users), $V=\{v_1,v_2,\hdots,v_n\}$, with $m$ edges that prescribe the feasible communication among the system entities. We have adopted a tree of depth 3 in our experiments: the global controller vertex $v_1$ resides at the root, local controllers 
reside at the first level, switches (2 switches per local controller) 
reside at the second level and hosts (users) lie at the third; the number of users is taken variable. In all test cases, we chose packet flows where the source and destination nodes are sampled uniformly at random.


We have studied four scenarios via altering a subset of parameters in our model while fixing others, as illustrated in Table \ref{tab1}: ``Requests'' represents the total number of served requests, ``Controllers'' refers to the number of local controllers in our topology, ``Users'' is the number of users for each switch, and ``Time'' denotes the accumulation of configuration time and test time.   
\begin{table}[h]
	\centering
	\caption{Experimental scenarios.}
	\label{tab1}
	\begin{tabular}{|c|c|c|c|c|}
		\hline
		\rowcolor[HTML]{BBDAFF} 
		\textbf{NO}                        & \textbf{\begin{tabular}{c} Requests\end{tabular}} & \textbf{Controllers} & \textbf{Users} & \textbf{Time} \\ \hline		
		\cellcolor[HTML]{BBDAFF}\textbf{Sc.1} & 10,000                                                            & Changed                        & 8                  & Tested              \\ \hline
		\cellcolor[HTML]{BBDAFF}\textbf{Sc.2} & 10,000                                                           & 8                       & Changed                  & Tested                   \\ \hline
		\cellcolor[HTML]{BBDAFF}\textbf{Sc.3} & Tested                                                            & 8                  & 8                  & Changed                   \\ \hline
		\cellcolor[HTML]{BBDAFF}\textbf{Sc.4} & Tested                                                            & Changed                  & Changed                  & Tested                   \\ \hline
	\end{tabular}
\end{table} 

\begin{figure}[h]
	\centering
	\includegraphics[width=0.45\textwidth]{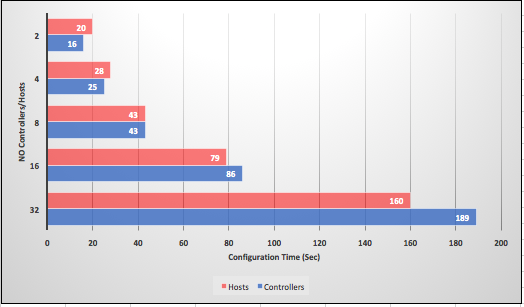}
	\caption{Configuration time for variable network size (scenarios 1\&2).}
	\label{fig:plot1}
\end{figure}

Figure \ref{fig:plot1} demonstrates the effect of varying the number of local controllers and users on the configuration time for the first two scenarios. We use blue bars to show the required time for variable number of controllers with the hosts per switch being fixed to 8 (scenario 1).  
We use red bars to illustrate the time needed for variable number of hosts per switch with a fixed number of 8 local controllers (scenario 2).  
It was observed that increasing the number of controllers requires more configuration time compared to increasing the number of users in the system for larger numbers,  
whereas the opposite was observed for smaller numbers. 
Additionally, a balanced time was reported when the number of controllers equals the number of hosts per switch (both equal to 8). 

\begin{figure}[h]
	\centering
	\includegraphics[width=0.45\textwidth]{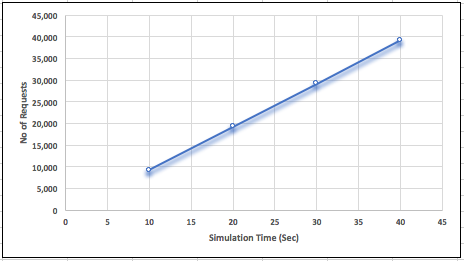}
	\caption{Number of requests served over simulation time (scenario 3).}
	\label{fig:plot2}
\end{figure}

Figure \ref{fig:plot2} shows the total number of requests served by our system for 8 local controllers and 8 hosts-per switch over several time periods. It is noted that, by design, the system gives an equally likely execution time for each packet across several time periods so that the number of requests served increases linearly with time.    

\begin{figure}[h]
	\centering
	\includegraphics[width=0.45\textwidth]{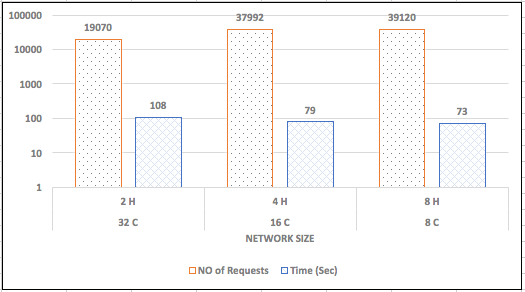}
	\caption{Number of requests served \& test turation time over variable network configurations (scenario 4).}
	\label{fig:plot3}
\end{figure}

Figure \ref{fig:plot3} illustrates the total number of served requests and simulation test time across three different network configurations for a fixed total number of users (which equals the product Number of Local Controllers * Number of Switches per Local Controller * Number of Hosts per Switch). Observe that 8 local controllers with 8 hosts/switch serve a little bit more than 16 controllers and 4 hosts with less amount of time. This reveals the benefit of obtaining optimal 
performance with minimum configuration cost for balanced topologies.  


\section{Conclusions}
\label{sec:con}

We have proposed a software-defined architecture for Cyberphysical Systems and IoT applications. We have specified the main requirements for different IoT applications in terms of performance, security, quality-of-service and real-time operation. We have demonstrated how the proposed model exploits the computational power of a great number of system components (coordinators, controllers, sensors, and portable devices) to control systems in a scalable and flexible way while prioritizing cyber security. All components are implemented as software defined nodes inside the middleware layer, and control and information flow in both top-bottom and bottom-up fashions. 
Finally, we have built a simulation testbed tool in Python to measure the performance of the proposed model, and ran extensive experiments that reveal the main benefits of the proposed model.

\bibliographystyle{spmpsci}      

\bibliography{references}


\end{document}